\documentclass[aps,twocolumn,superscriptaddress,floatfix,showpacs] {revtex4}

\usepackage{epsfig}
\usepackage{graphicx}
\usepackage{dcolumn}
\usepackage{bm}

\begin{document}

\title{X-ray spectra and electronic structure of FeAs superconductors}

\author{E.Z.~Kurmaev}
\affiliation{Institute of Metal Physics, Russian Academy of Sciences-Ural Division, 620219 Yekaterinburg, Russia}

\author{R.G.~Wilks}
\affiliation{Department of Physics and Engineering Physics, University of Saskatchewan, 116 Science Place Saskatoon, Saskatchewan S7N 5E2, Canada}

\author{A.~Moewes}
\affiliation{Department of Physics and Engineering Physics, University of Saskatchewan, 116 Science Place Saskatoon, Saskatchewan S7N 5E2, Canada}

\author{N.A.~Skorikov}
\affiliation{Institute of Metal Physics, Russian Academy of Sciences-Ural Division, 620219 Yekaterinburg, Russia}

\author{Yu.A.~Izyumov}
\affiliation{Institute of Metal Physics, Russian Academy of Sciences-Ural Division, 620219 Yekaterinburg, Russia}

\author{L.D. Finkelstein}
\affiliation{Institute of Metal Physics, Russian Academy of Sciences-Ural Division, 620219 Yekaterinburg, Russia}

\author{R.H.~Li}
\affiliation{Hefei National Laboratory for Physical Science at Microscale and Department of Physics, University of Science and Technology of China, Hefei, Anhui 230026, People's Republic of China}

\author{X.H.~Chen}
\affiliation{Hefei National Laboratory for Physical Science at Microscale and Department of Physics, University of Science and Technology of China, Hefei, Anhui 230026, People's Republic of China}

\date{\today}

\begin{abstract}   
The densities of the valence and conduction band electronic states of the newly discovered layered superconductors LaOFeAs, LaO$_{0.87}$F$_{0.13}$FeAs (T$_c$=26 K), SmO$_{0.95}$F$_{0.05}$FeAs and SmO$_{0.85}$F$_{0.15}$FeAs (T$_c$=43 K) are studied using soft X-ray absorption and emission spectroscopy combined with FP LAPW calculations of LaOFeAs, LaO$_{0.875}$FeAs and LaO$_{0.875}$F$_{0.125}$FeAs. The Fe $3d$-states are localized in energy near the top of the valence band and are partially hybridized with p-type O ($2p$), As ($4p$), and La ($6p$) states approximately 3~eV below the Fermi energy. The Fe $L_3$ X-ray emission spectra do not show any features that would indicate the presence of the low Hubbard 3d-band or the quasiparticle peak that were predicted by the DMFT analysis of LaOFeAs. We can conclude that the LaOFeAs-type compounds do not represent strongly correlated systems. When either oxygen vacancies or fluorine dopants are included in numerical electronic structure calculations the width of O $2p$-band decreases, but the distribution of Fe $3d$-states is largely unaffected.
\end{abstract}


\pacs{ }
\maketitle

\section*{Introduction.}
The discovery of high-T$_c$ superconductivity in Ln(O$_{1-x}$F$_x$)FeAs (Ln= La, Ce, Pr, Nd, Sm, Gd) with T$_c$ $\approx$ 25-55 K \cite{1,2,3,4,5,6,6a} and high upper critical fields H$_{c2}$  up to 65~Tesla \cite{7} has spurred  a great interest to class of layered oxypnictides (LnO)(MPn) (where M=$3d$ metals, Pn=P, As). The observation of high-T$_c$ superconductivity in these systems is surprising, because the magnetic $3d$ elements in such a material would be expected to produce magnetic moments that give rise to a long-range ferromagnetic order which would be unfavourable for superconductivity with singlet pairing.

\begin{figure}[htb]
\centering
\includegraphics[width=0.7\linewidth]{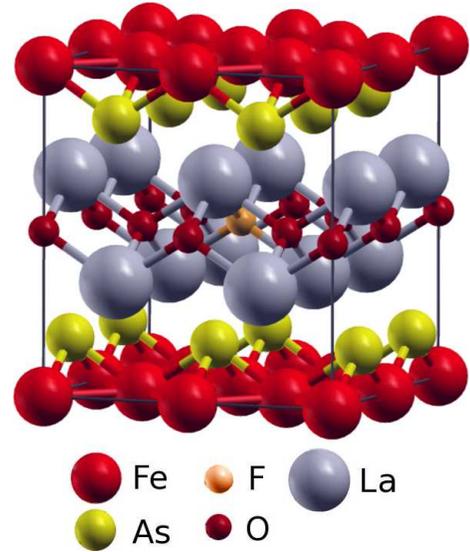}
\caption{Crystal structure of LaO$_{0.875}$F$_{0.125}$FeAs.}
\label{fig1}
\end{figure}

The structure of these superconductors (see Fig.~\ref{fig1}) is analogous to that which is seen in high-T$_c$ cuprates, with an FeAs plane in place of an CuO based plane. These new superconductors have a layered structure in which the role of the copper oxygen plane in the high-T$_c$ cuprates is played by the FeAs plane. The Fe ions are arranged in a simple square lattice, and one can extend the analogy with the cuprate materials to include the supposition that the superconducting mechanism includes electron hopping in the Fe lattice and doping from the nearby oxide charge reservoir layer. On the other hand, one can notice some differences: undoped LaOFeAs has a metallic conductivity at room temperature which is opposite to cuprates -- correlated insulators. 

To understand the electronic structure properties of the new Fe-based superconductors and the interplay between the crystal structure and magnetic states, first-principles density functional theory (DFT) calculations in the local density approximation (LDA) have been performed for the undopedand fluorine doped LaOFeAs \cite{8,9,10,11,12,13,14}. Most of the LDA calculations show a high density of Fe-$3d$-states at the Fermi level, whereas the DMFT calculation in the paramagnetic phase, including strong-correlation effects beyond LDA, shows that for U = 4~eV, a large amount of that spectral weight is shifted away from the Fermi level, and the undoped system has a bad metallic behavior \cite{15}. To date, only one experimental study of the electronic structure of LaOFeAs has been published \cite{16}; it presented angle-integrated photoemission spectra showing a a sharp feature very close to the Fermi energy and a relative flat distribution of the density of states between 0.5~eV and 3~eV binding energy. In the present study, we examine the resonant X-ray emission spectra (RXES) at Fe $L$-edge and non-resonant oxygen and fluorine $K$-emission and absorption spectra of undoped and fluorine doped LnOFeAs (Ln=La, Sm), providing an experimental indication of the distribution of Fe $3d$, O $2p$ and F $2p$ occupied and vacant electronic states. 

\section{Experimental and Computational Details.}
Polycrystalline samples with nominal composition SmO$_{1-x}$F$_x$FeAs (x=0.05, 0.15)  were synthesized by a conventional solid state reaction using high purity SmAs, SmF$_3$, Fe and Fe$_2$O$_3$ as starting materials. SmAs was obtained by reacting Sm chips and As pieces at 600$^\circ$~C for 3 hours and then 900$^\circ$~C for 5 hours. The raw materials were thoroughly grounded and pressed into pellets. The pellets were wrapped into Ta foil and sealed in an evacuated quartz tube. They were then annealed at 1160$^\circ$~C for 40 hours. With the exception of the annealing, the sample preparation process was carried out in a glove box with a high-purity argon atmosphere. The  synthesis of the LaO$_{1-x}$F$_x$FeAs (x=0, 0.13), is identical except for the substitution of LaAs and LaF$_3$ in the starting materials. For a more complete description of the sample preparation see Ref.~\onlinecite{4}.

The soft X-ray absorption and emission measurements of the fluorine doped LaOFeAs and SmOFeAs were performed at the soft X-ray fluorescence endstation at Beamline 8.0.1 of the Advanced Light Source at Lawrence Berkeley National Laboratory \cite{18}. The endstation uses a Rowland circle geometry X-ray spectrometer with spherical gratings and an area sensitive multichannel detector. We have measured the resonant and non-resonant Fe $L_{2,3}$ ($3d4s \to 2p$ transition) and non-resonant F $K\alpha$ and O $K\alpha$  ($2p \to 1s$ transition) X-ray emission spectra (XES). The instrument resolution for F $K\alpha$ and Fe $L_{2,3}$ X-ray emission spectra was 0.8~eV and for O $K\alpha$ XES -- 0.5~eV. X-ray absorption spectra (XAS) were measured in the total fluorescence mode with a resolwing power  $E/\Delta E$=5000. All spectra were normalized to the incident photon current using a clean gold mesh in front of the sample to correct for intensity fluctuations in the photon beam. The excitations for the RXES measurements were determined from the XAS spectra; the chosen energies corresponded to the location of the $L_3$ and $L_2$ thresholds as well as one energy well above resonance. 

All band structure calculations were performed within the full-potential augment plane-wave method as implemented in WIEN2k code \cite{19}. For the exchange-correlation potential we used gradient approximation \cite{20} in the Perdew-Burke-Ernzerhof variant. The Brillouin zone integrations were performed with a $12\times 12\times 5$ special point grid and $R_{MT}^{min}K_{max}$=7 (the product of the smallest of the atomic sphere radii $R_{MT}$ and the plane wave cutoff parameter $K_{max}$) was used for the expansion of the basis set. The experimentally determined lattice parameters as well as internal positions of LaOFeAs (a=4.03007~\AA, c=8.7368~\AA) \cite{3} were used. The spheres radii were chosen $R_{Fe}$=2.40, $R_{As}$=2.13, $R_{La}$=2.35, $R_{O}$=2.09 and $R_{F}$=2.09~a.u. They were chosen in such a way that the spheres are nearly touching. To calculate band structure of LaO$_{7/8}$F$_{1/8}$FeAs we constructed a $2a\times 2a\times c$ supercell (where $a$ and $c$ are experimentally determined lattice parameters of LaO$_{0.92}$F$_{0.08}$FeAs \cite{21}) in which one of oxygen atoms was replaced by fluorine atom. The resulting unit cell has a lower symmetry than initial unit cell of LaOFeAs: space group P-4m2 (115). Calculations of the electronic structure of LaO$_{7/8}$FeAs were performed for the same crystal structure as for LaO$_{0.92}$F$_{0.08}$FeAs but with a vacancy in place of fluorine atom. The inclusion of this vacancy in the supercell does not change its symmetry from that of the previous supercell.  For the comparison with the experimental spectra the obtained DOS curves were broadened with Lorentz functions of width 0.3 eV.

\section{Results and discussion.}
\subsection{Undoped LaOFeAs.}
\begin{figure}
\centering
\includegraphics[width=0.6\linewidth]{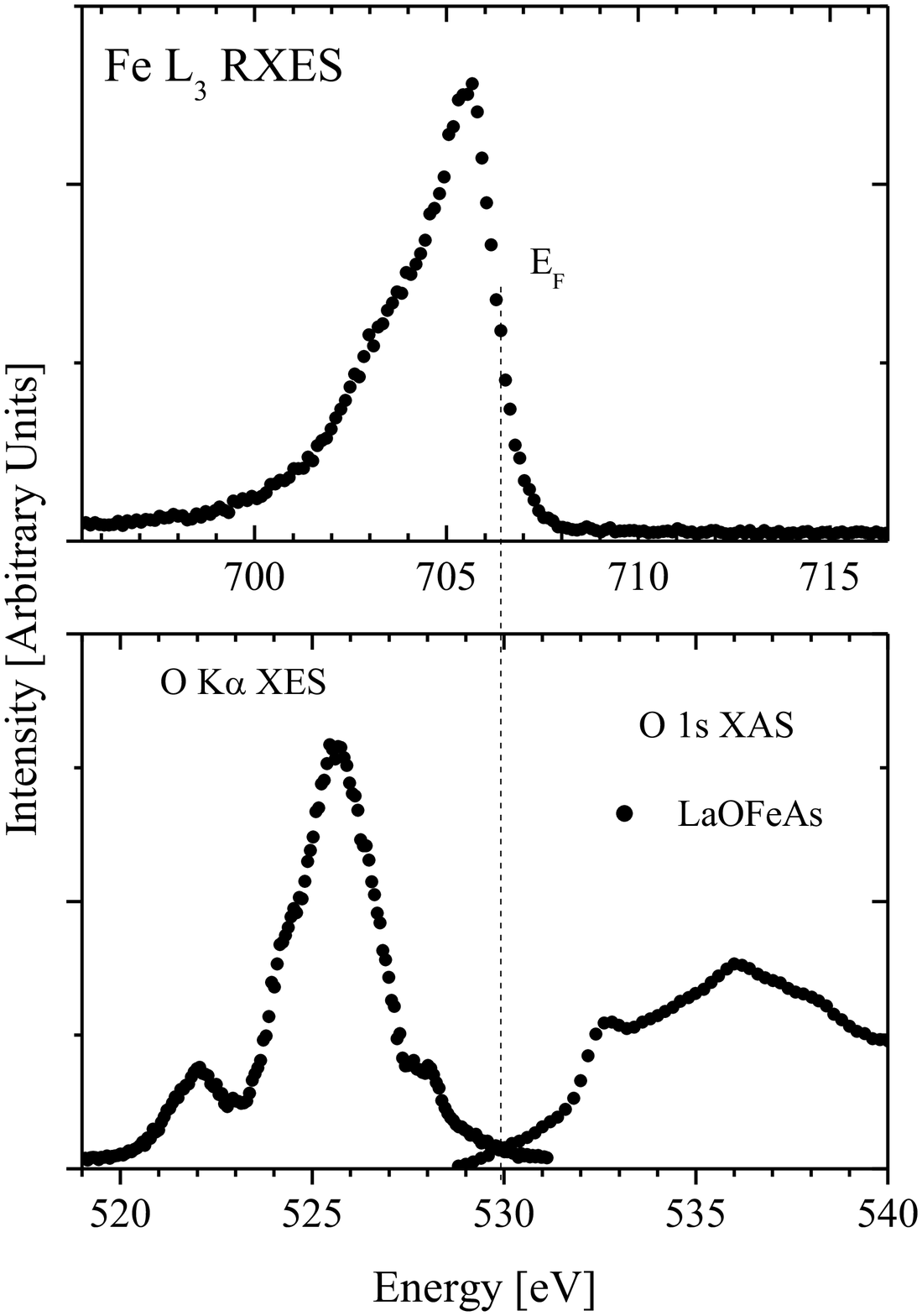}
\caption{Soft X-ray spectra of constituents of LaOFeAs: Oxygen {\it K}-emission and absorption spectra of LaOFeAs (lower panel) and resonantly excited Fe $L_3$ X-ray emission spectrum.}
\label{fig2}
\end{figure}

\begin{figure}
\centering
\includegraphics[width=0.6\linewidth]{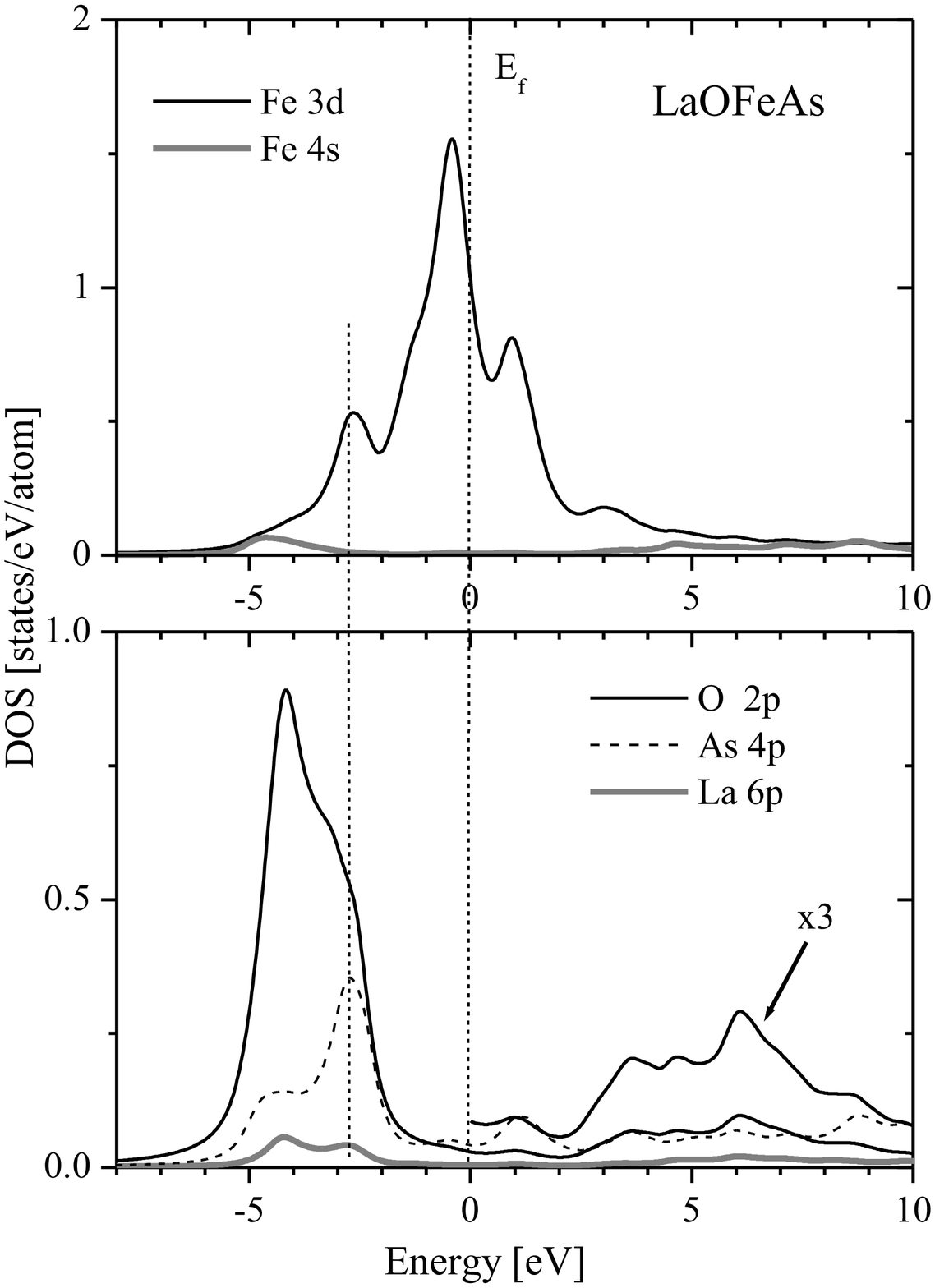}
\caption{Calculated partial DOSes of LaOFeAs.}
\label{fig3}
\end{figure}

The oxygen K-emission and absorption spectra of undoped LaOFeAs are presented in Fig.~\ref{fig2}. 
The calculated partial density of states (DOS) of LaOFeAS are  are shown in Fig.~\ref{fig3}.
The comparison of experimental oxygen $K$-emission and absorption spectra and calculated O $2p$ DOS shows a good correspondence. As seen, in the vicinity of the Fermi level E$_\mathrm{f}$ (determined as an intersection of O $K\alpha$ XES and O $1s$ XAS) the occupied O $2p$-states are negligible and mostly are concentrated at the middle of the valence band. O $1s$ XAS shows features at $\sim$0.7, 2.6 and 5.9~eV above the Fermi level which can be related to calculated O $2p$-peaks in the conduction band at 0.9, 3.3 and 6.0~eV. 

In the absence of measurements of XPS Fe $2p$-binding energies of LaOFeAs we have determined the Fermi level on the Fe $L_3$ XES as high-energy edge cut off and using this procedure found that the main spectral weight of Fe $L_3$ XES is concentrated at $\sim$0.9~eV below E$_\mathrm{f}$ which correlates with metallic behavior of LaOFeAs and a low-energy shoulder at around 3.0~eV below E$_\mathrm{f}$. This experimental distribution of Fe $3d$-states corresponds to our LDA electronic structure calculation (Fig.~\ref{fig3}) and is in contradiction with calculated DMFT density of states \cite{15} according to which the main weight of Fe $3d$-states is shifted to approximately 4~eV below the Ef forming low Hubbard $3d$-band and quasiparticle peak near the Fermi level. The recent LDA+DMFT(QMC) study of LaOFeAs also show that LaOFeAs is far from the strongly correlated regime \cite{22}. 

\subsection{Oxygen deficient and fluorine doped LaOFeAs and SmOFeAs}
\begin{figure}[htb]
\centering
\includegraphics[width=0.6\linewidth]{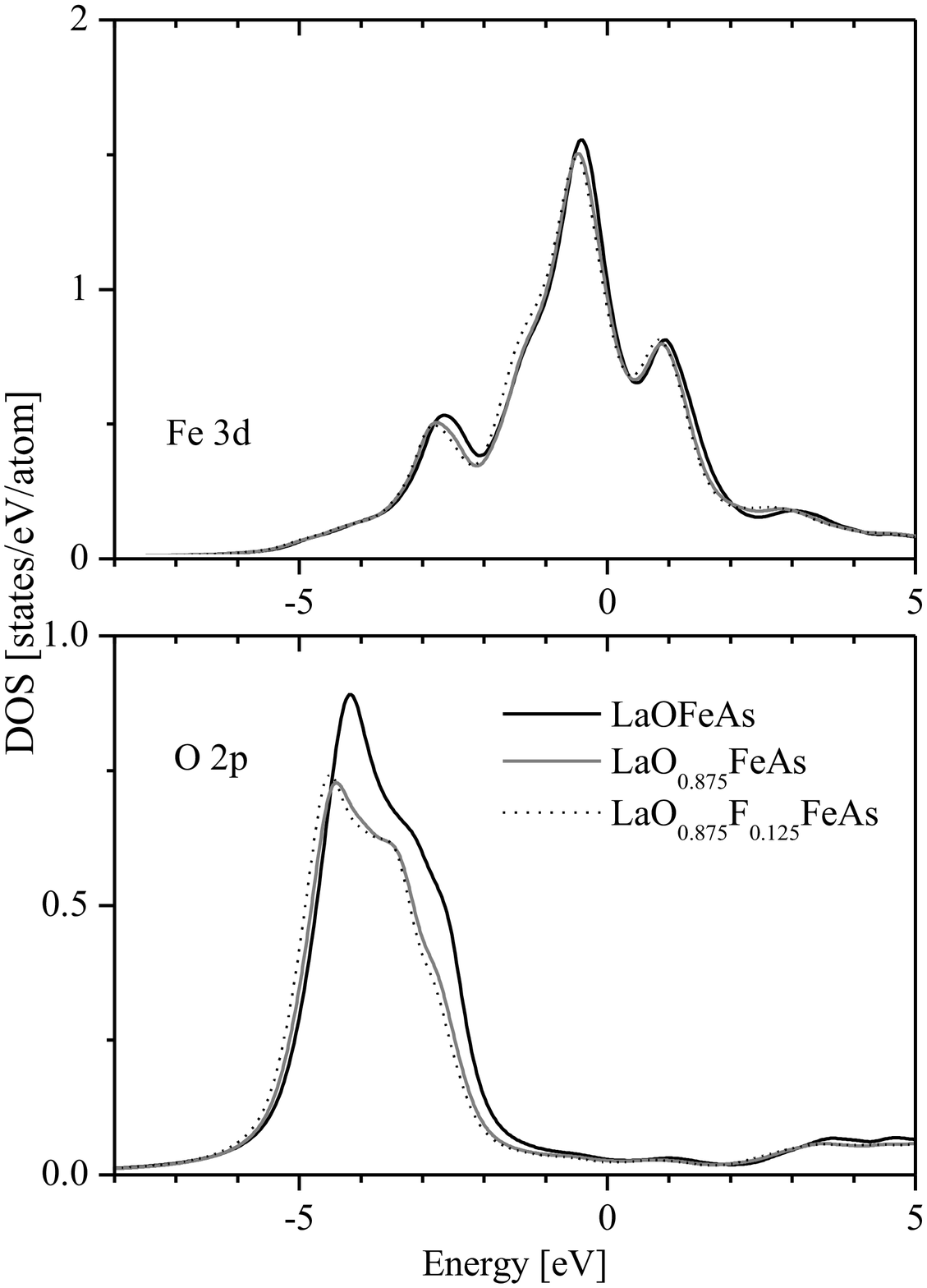}
\caption{The comparison of calculated partial Fe $3d$ and O $2p$ DOSes of LaOFeAs La, LaO$_{0.875}$FeAs and LaO$_{0.875}$F$_{0.125}$FeAs.}
\label{fig4}
\end{figure}

The recent observation of high-T$_c$ superconductivity with T$_c$=31-55~K for undoped oxygen deficient LnO$_{0.85}$FeAs (Ln=La, Ce, Pr, Nd, Sm) \cite{24} suggest that the effect of fluorine doping in these samples may be related to the oxygen vacancy effect. The calculated electronic structures of LaOFeAs, nonstoichiometric LaO$_{0.875}$FeA, and fluorine doped LaO$_{0.875}$F$_{0.125}$FeAs are shown in Fig.~\ref{fig4}. 
\begin{figure}[htb]
\centering
\includegraphics[width=0.6\linewidth]{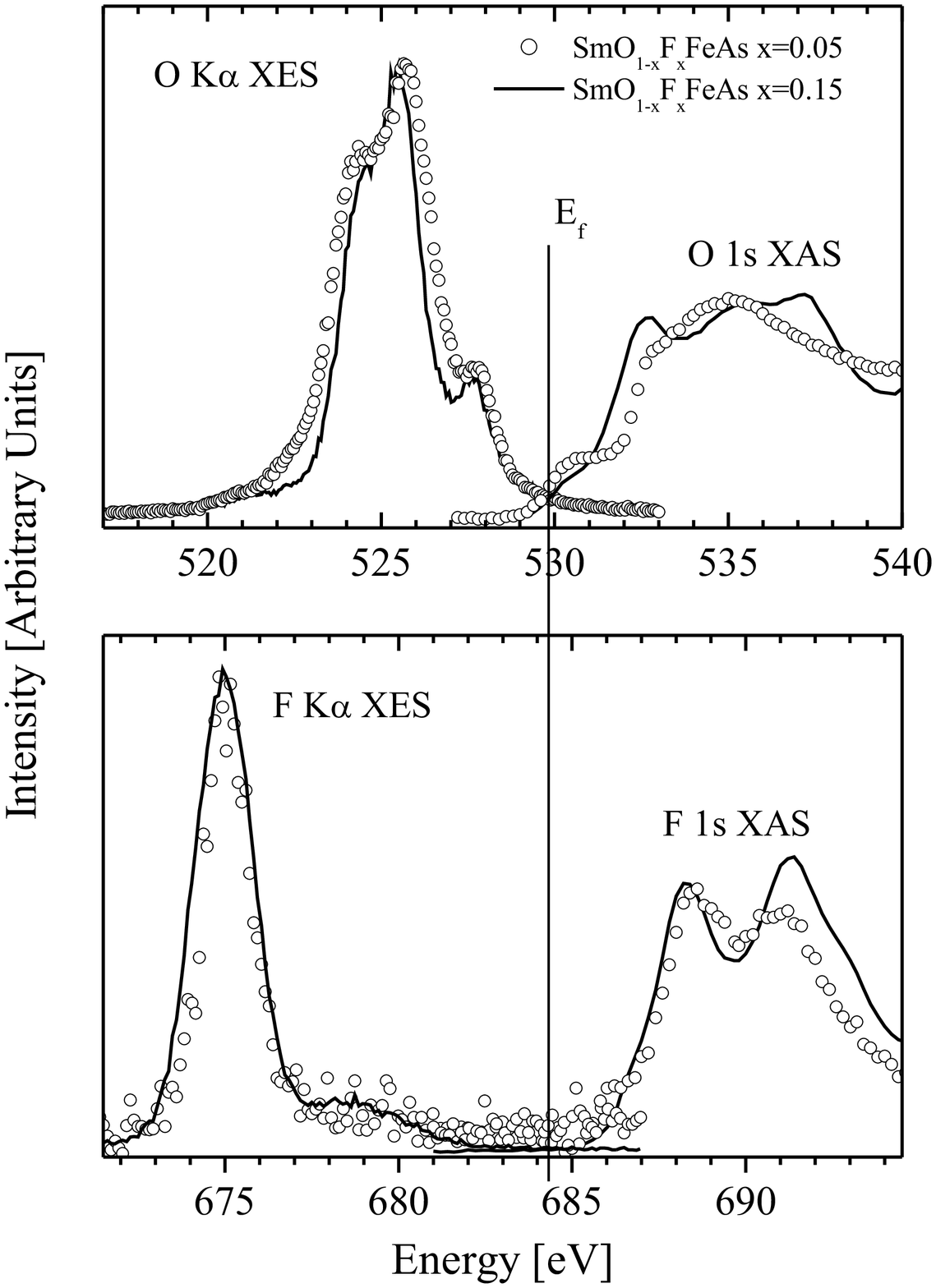}
\caption{Fluorine (lower panel) and oxygen (upper panel) K-emission and absorption spectra of  Sm$_{1-x}$F$_x$OFeAs (x=0.05, 0.15). }
\label{fig5}
\end{figure}
The O 2p-band of oxygen deficient and fluorine doped LaOFeAs is narrower than undoped one, whereas the Fe $3d$-states remain essentially unchanged. We need to point out that O $2p$ DOS for the same level of oxygen deficiency or fluorine doping (12.5\%) is found to be the same. Assuming that the carriers in the FeAs reservoir layer are provided by the oxygen vacancies, it can be inferred that the substitution of fluorine in the oxygen sites produces the same effect which is seen in reduction of O $2p$ band width. The reduction of calculated O $2p$ band width of fluorine doped LaOFeAs with respect to undoped compound is confirmed by direct measurements of the full width at half maximum (FWHM) of the O $K\alpha$ XES for fluorine-doped SmOFeAs (Fig.~\ref{fig5}). The resonant excitation of Fe $L_{2,3}$ XES at $L_3$-threshold selectively excites the Fe $L_3$- XES, eliminating the contribution from the Fe $L_2$ XES and allowing for a more clear evaluation of the effects of fluorine-doping on the Fe $3d$-states to be made. In agreement with the calculations of the electronic structure, the Fe $L_3$ emission of the LaOFeAs and SmOFeAs compounds (Fig.~\ref{fig7}) are unchanged as a result of the doping.

\begin{figure}[htb]
\centering
\includegraphics[width=0.65\linewidth]{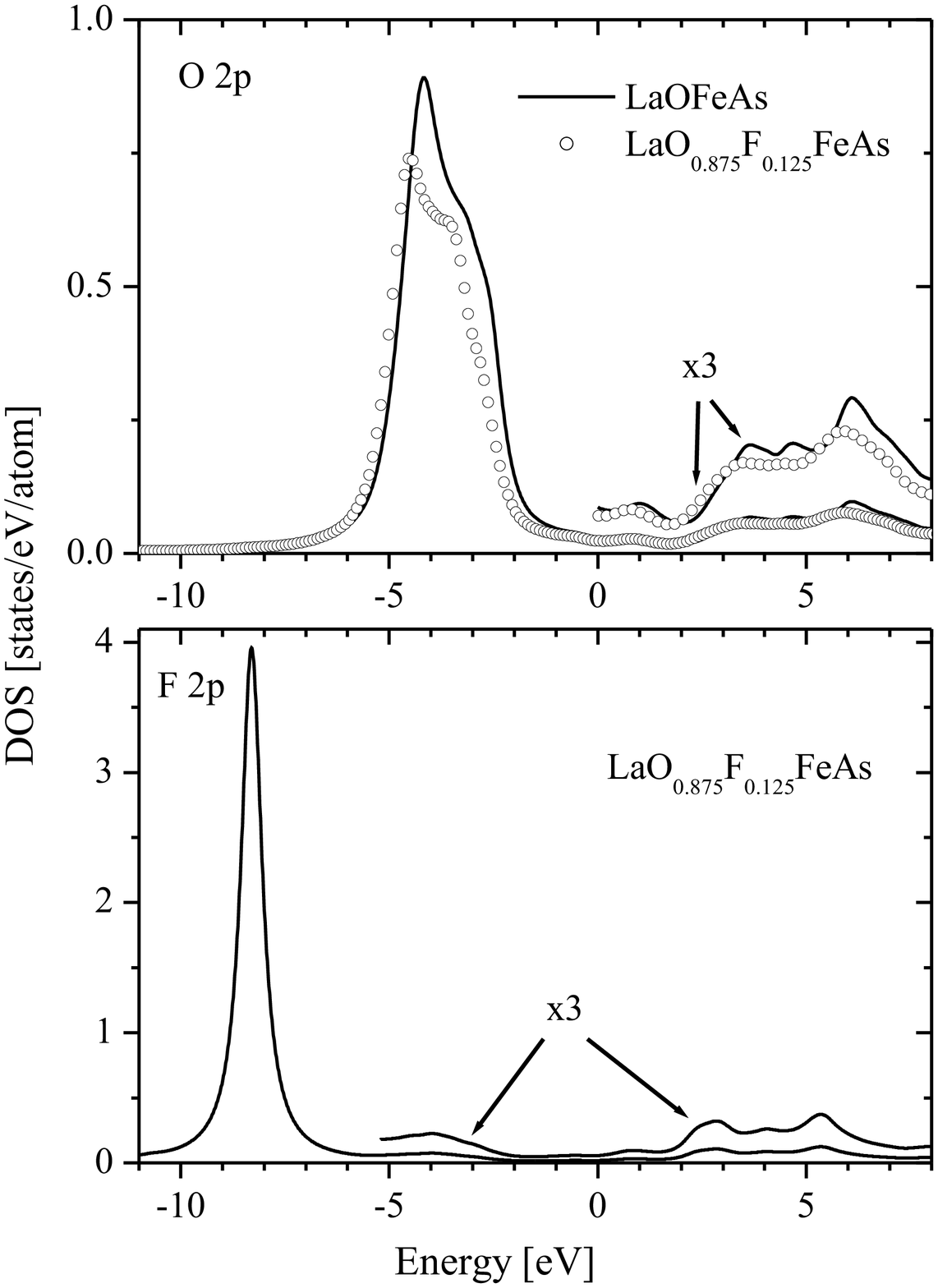}
\caption{Calculated fluorine (lower panel) and oxygen (upper panel) $2p$ DOS of LaO$_{0.875}$F$_{0.125}$FeAs.}
\label{fig6}
\end{figure}
\begin{figure}[htb]
\centering
\includegraphics[width=0.65\linewidth]{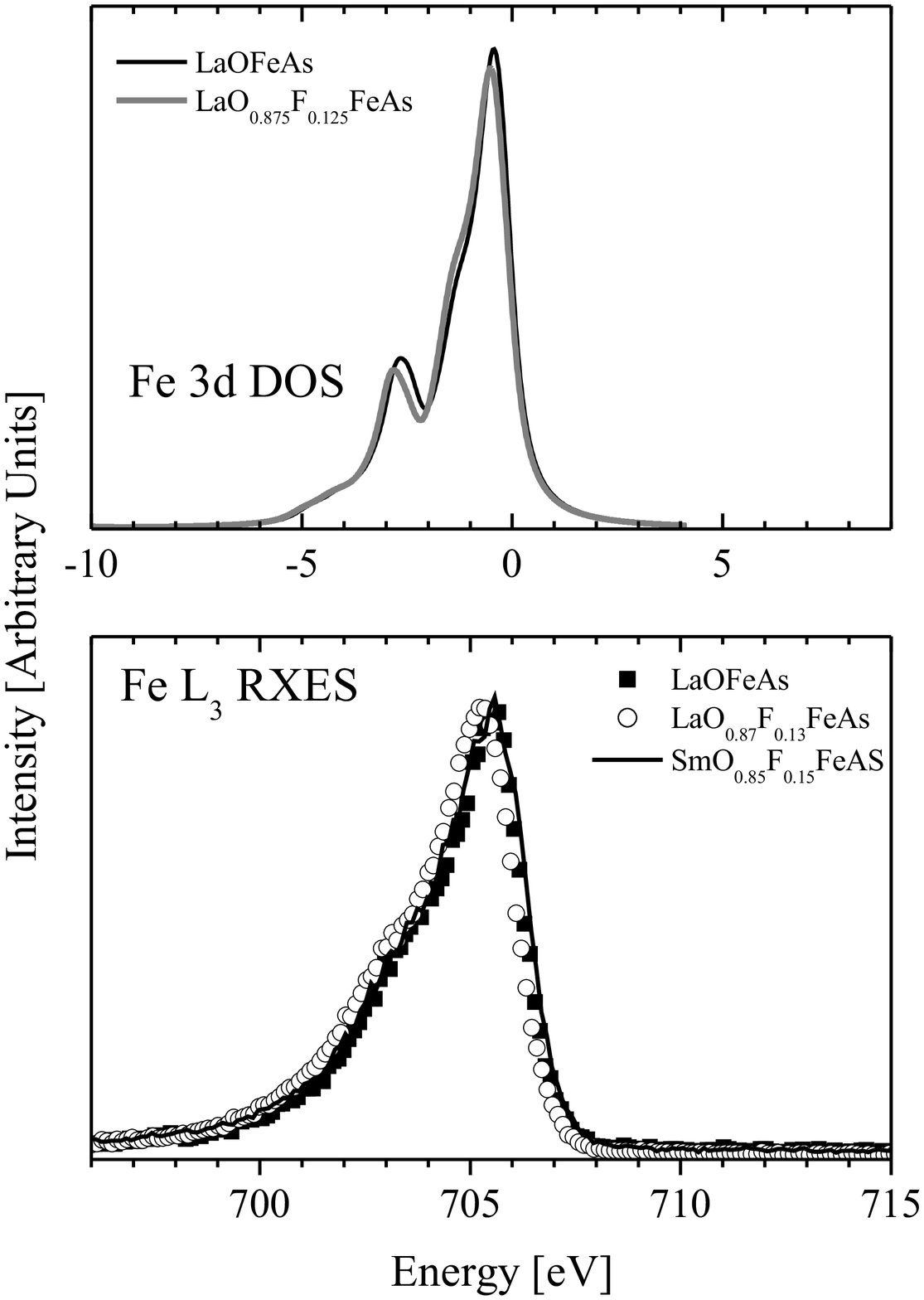}
\caption{The comparison of Fe $L_3$ RXES (lower panel) and calculated Fe $3d$ DOS of undoped and fluorine doped FeAs superconductors.}
\label{fig7}
\end{figure} 
The experimental F $K\alpha$ XES (Fig.~\ref{fig5}) and calculated F $2p$ DOS (Fig.~\ref{fig6}) show that F $2p$-states are located about 9.1 and 8.2~eV below the Fermi level, respectively, and not mixed with Fe $3d$ and O $2p$-states. Therefore the role of F $2p$-states seems to provide the necessary amount of electron carriers to FeAs-layer. In agreement with the calculations of the electronic structure, the Fe $L_3$ emission of the LaOFeAs and SmOFeAs compounds (Fig.~\ref{fig7}) are unchanged as a result of the doping.

\section*{Conclusion.}
In the present paper we have studied the electronic structure of undoped, oxygen deficient and fluorine doped FeAs superconduictors by RIXS measurements and \textit{ab-initio} band structure calculations. The resonantly excited Fe $L_3$ X-ray emission spectra show that occupied Fe $3d$-DOS has two-peaks structure with the main peak formed by pure Fe $3d$-states and located at $\sim$1.0~eV below the Fermi level and low-energy subband at $\sim$3.0~eV with respect to E$_\mathrm{f}$ which is due to hybridization of Fe $3d$-states with O $2p$ (As $4p$ - La $6p$) states. These results are in a good agreement with our FP LAPW electronic structure calculations. We have not found the features predicted by DMFT calculations: low Hubbard $3d$-band below E$_\mathrm{f}$ for about $\sim$4~eV and quasiparticle peak at the Fermi level and therefore we conclude that the FeAs-type superconductors are not strongly correlated systems. Electronic structure calculations performed for LaO$_{0.85}$FeAs and LaO$_{0.85}$F$_{0.15}$FeAs show the identical results for the same level of oxygen deficiency or fluorine doping: unchanged Fe $3d$-DOS and reduced band width of O $2p$-band. The reduction of width of O $2p$-band under F-doping is confirmed by direct measurements of O $K\alpha$ XES of SmO$_{0.85}$F$_{0.15}$FeAs. Therefore the role of oxygen vacancies and fluorine doping seems to be the same - to form oxide charge reservoir layer and supply FeAs planes by carriers which are necessary for occurrence of the same superconducting properties. This situation reminds fluorine doping in cuprates which gives the similar results. 

We acknowledge support from the RAS Program (project 01.2.006 13395) with partial support of the Research Council of the President of the Russian Federation (Grants NSH-1929.2008.2 and NSH-1941.2008.2), the Russian Science Foundation for Basic Research (Project 08-02-00148), the Natural Sciences and Engineering Research Council of Canada (NSERC), and the Canada Research Chair program.

\end{document}